# Reply to "Failure to replicate long-range tunable attractions in colloidal system"

Bo Li*[1], Feng Wang[1], Di Zhou[1], Yi Peng[1], Ran Ni[2], and Yilong Han**[1],
* bliad@connect.ust.hk
** yilong@ust.hk
1. Department of Physics, Hong Kong University of Science and Technology, Clear Water Bay, Hong Kong, China
2. School of Chemical and Biomedical Engineering, Nanyang Technological University, Singapore

An arxiv paper, ref. [1] by Cao et al., claimed that the tunable attraction reported in our ref. [2] could not be detected. Ref. [1] was submitted to Nature in Apr. 2016 as a Comment on our ref. [2]. Our reply in May 2016 responded to ref. [1] and was reviewed by the editor of Nature and an external referee. Ref. [1] was rejected by Nature in Aug. 2016. We provide our reply, which answers all of the criticisms in ref. [1], in Appendix I. Below is a brief reply to the main criticisms.

There are two main criticisms to Ref. [1] (others are irrelevant, or misleading, see Appendix I): 1) the $g(r)$ in the Supplementary Information (SI) of ref. [2] could not produce $u(r)$ in the main text, and 2) Cao could not experimentally reproduce the tunable attraction at a low colloid density. In response to the first question, the issue arose because our reported $g(r)$ curves were smoothed, but our reported $u(r)$ curves were obtained from the raw $g(r)$. The unsmoothed $g(r)$ and $u(r)$ are provided in Fig. A1 in Appendix II. Responding to the second question, this observation is likely due to inappropriate sample preparation, e.g. using excessive dye or objective oil, causing the objective heater to heat the entire sample area without the dye pumping effect. Prof. Ziren Wang's group easily reproduced the tunable attraction in Sept. 2016 (see Fig. A2 in Appendix II).

We found that the tunable attraction arises from the dye pumping effect. The $(2 \text{ cm})^2$ sample cell was partially heated using an objective heater through a thermal contact of a roughly 1 $cm^2$ circular area of the immersion oil of the 100x objective. The temperature gradient at the perimeter of the 1 $cm^2$ heated area induced the thermophoretic pumping of the dye from the ambient area into the heated area, resulting in a higher dye concentration and stronger attraction in the heated area. At each temperature increment, the sample reached a steady state within 30 min. We recorded the particles' motion after one-hour equilibration at each temperature increment. The $(0.1 \text{ mm})^2$ field of view is at the center of the 1 $cm^2$ heated area, and thus the temperature was uniform in the field of view.

The key evidence of tunable attraction is shown in Fig. R2 of Appendix I, which depicts the same field of view without moving the objective. Cao et al. questioned whether Li shifted the field of view during the experiment. This is currently the main issue in dispute. To demonstrate that the tunable attraction is genuinely observed without shifting the field of view, here we provide videos from two experiments: one video set with stuck particles in the field of view and one continuous full-time video at Youtube.

Video 1: The continuous full-time video. t = 0 min, T = 24.0°C; t = 2 min, T = 26.0°C; t = 8 min, T = 28.0°C; t = 10 min, T = 30.0°C. Their images are shown in Fig. 1. The room temperature was fixed at 23.0°C. 100x real time.
https://www.youtube.com/watch?v=wwlGJKQs_CM

Video Set 2: Videos with several particles stuck on the substrate. v0, T=32.0°C; v1, T=30.0°C; v2, T=28.0°C; v3, T=26.0°C. The room temperature was fixed at 24.0°C. Their raw images are shown in Fig. A5. The positions of stuck particles did not move in different videos, demonstrating that the field of view was not shifted.
https://www.youtube.com/watch?v=MjHNqRXp2vU
https://www.youtube.com/watch?v=4Ini3Y4xJAg
https://www.youtube.com/watch?v=r8dsKfV4BDA
https://www.youtube.com/watch?v=59W-5aMpmi4

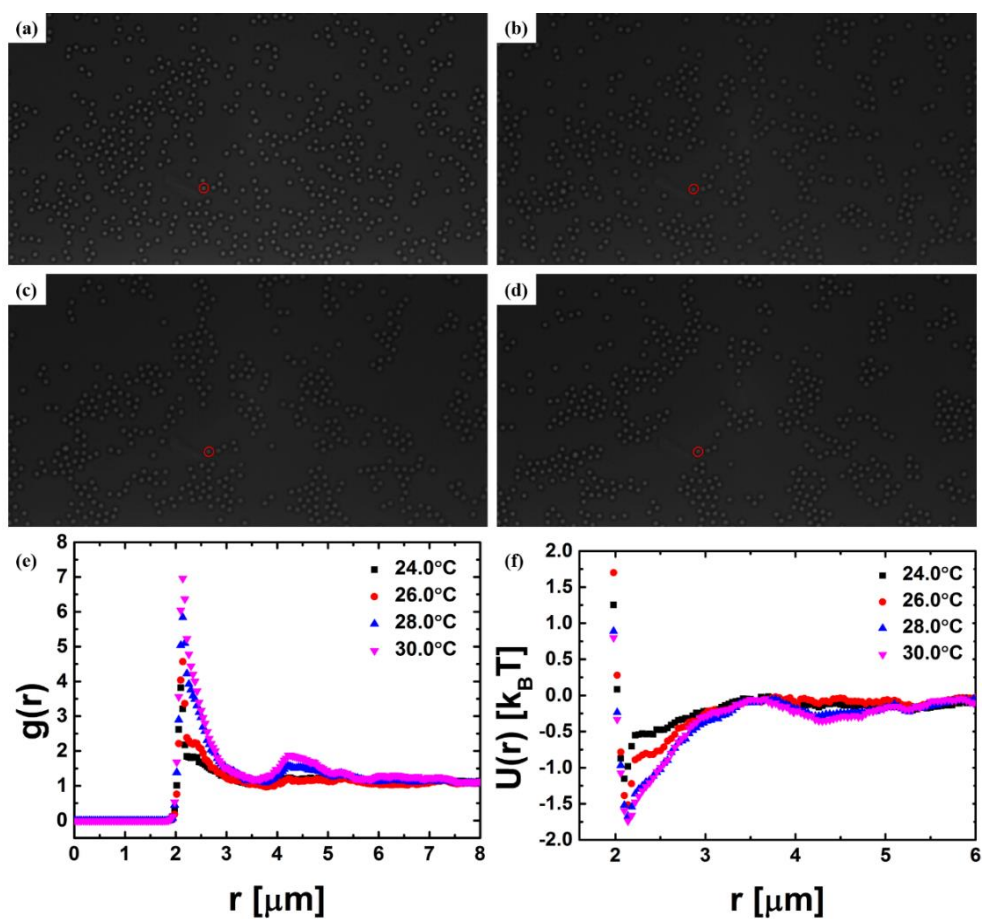

Fig. 1. The raw images at (a) 24.0°C, (b) 26.0°C, (c) 28.0°C and (d) 30.0°C from

movie 1. The cluster formation clearly demonstrates strong attraction at high temperatures. A field of view with several stuck particles (one of them is labeled in the red circle) was chosen to eliminate the risk of changing it. The colloid area in this sample has a elongated shape to maximize tunability. (e) The measured radial distribution functions. (f) The pair potentials derived from g(r) in (e).

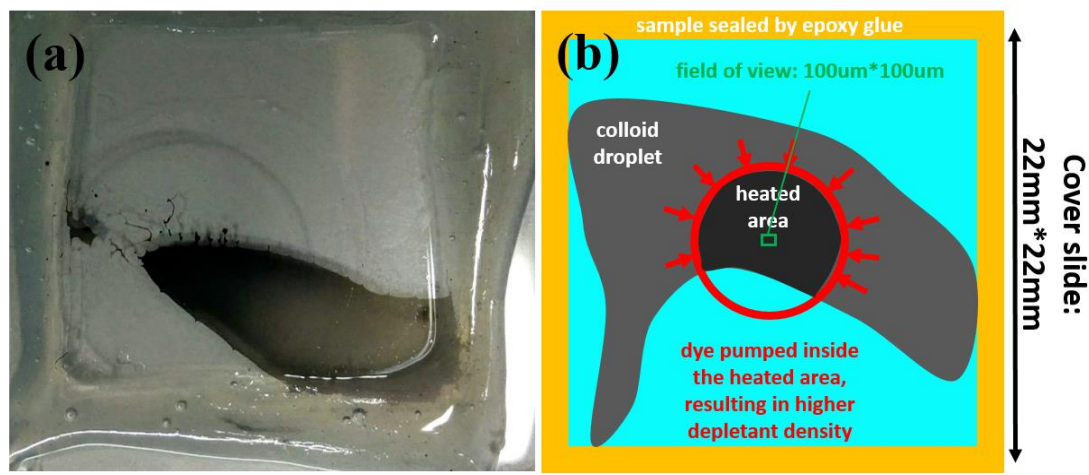

Fig. 2. (a) A elongated shaped colloid sample has high tunability, can form clusters and is visible to the naked eye. The tunability in ref. [1] was less strong because the colloid area was more circular. The heated area is darkened relative to the non-heated area due to the dye pumping effect. (b) A schematic illustration of the dye pumping in the sample. As far as we are aware, the dye pumping effect never occurred within Cao's sample in ref. [2] and subsequent debates. Since the crystal tend to grow from the colloid-wall and colloid-air interfaces, it is easier to observe crystals when a colloid-air interface is close to the field of view.

Other similar evidence of tunable attraction is provided in Appendix II.

The tunability of the attraction is low in a fresh mixture of the dye and the PMMA colloid. The tunability increases with time and saturates to a stable level after three months. An elongated colloidal droplet in the glass cell usually has higher tunability of attraction (Fig. 2) than a circular-shaped colloidal droplet because the smaller heated area causes the dye concentration to change more sensitively.

**References:**

[1] Xin Cao, Maijia Liao and Xiao Xiao, arxiv.1604.07723 (It is identical to Cao et al.'s Comment submitted to Nature in Apr. 2016.)
[2] Bo Li, Feng Wang, Di Zhou, Yi Peng, Ran Ni, and Yilong Han, Nature, 531, 485 (2016).

**Contribution of this manuscript:** Li performed all the experiments and data analysis. Han and Li wrote the manuscript. All coauthors discussed the results.

**Appendix I:** Reply to Cao et. al's Comment (i.e. ref.[1]) that we submitted to Nature in May 12$^{th}$, 2016.

We categorize all criticisms in the Comment into the following 4 questions for clarity. **Questions 1, 2B and 4 are new. Questions 2A and 3 have already been raised and addressed in the report to referees in Nov. 2015**; here we add recent new results and understandings.

1. **Why the radial distribution function g(r) at 21°C and 30°C in the published Supplementary Fig.S1 cannot produce the potential U(r) in Fig.1 of ref.1?**

It is because of the smoothing effect and the inaccurate area densities. U(r) attraction depth depends on the g(r) peak height and the area density. For the same g(r) peak height, a lower area density corresponds to a stronger attraction.

We plotted the smoothed g(r) in Fig.S1 in order to clearly show the 7 g(r) in one plot, but actually used raw g(r) to calculate U(r) then smoothed this U(r) as shown in Fig.1 of ref.1. Smoothing is a weighted average of several neighboring bin values, hence bins with g(r) > 1 near the peak are lowered toward 1 as shown in Fig.R1 (bins with g(r) < 1 at small r are increased toward 1). Consequently when Cao *et al.*'s used such smoothed g(r), they will get a weaker attraction in Comment Fig.1b and even unrealistic long-range repulsion in Comment Fig.1a. Different ways to get U(r) are shown in Fig.R1. Note that the symmetric errors in g(r) become asymmetric after they propagate to U(r) because the algorithm of g(r) → U(r) acts as a nonlinear function.

We regret that our Supplementary Information (SI) of ref.1 was a bit sloppy on the following two points: 1. We showed the smoothed g(r) which looks better than raw g(r), but makes readers cannot accurately reproduce U(r) at some of the temperatures. 2. As the input parameter in g(r) → U(r) algorithm, the area fractions were 15%, 16%, 16%, 16%, 14%, 14%, 13% at seven different temperatures, but we only reported a typical area fraction of 16% in the caption of Fig.S1.

Nevertheless, the trend of the increasing peak in raw g(r), i.e. increasing the U(r) depth, can be clearly seen. In fact, the precise curves of U(r) are not necessary because the g(r) → U(r) algorithm assumed that U(r) is pairwise additive (i.e. not just a function of r, but also depends on the position of neighboring particles). Pairwise additivity holds in dilute samples for our g(r) measurements, but not in dense crystal phase. Most atomic and colloidal interactions are non-pairwise additive at high densities, such as depletion attraction, Casmir-like attraction and screened-Coulomb interaction for colloids. g(r) and U(r) just qualitatively illustrated that the attraction is tunable, and cannot quantitatively reflect how the non-pairwise additive force changes in premelting crystals. The non-pairwise additivity has been pointed out in our published SI.

In addition, the small polydispersity and image-tracking noises broaden the measured g(r) peak and make the peak height lower than the true value, leading to an underestimate of U(r) attraction depth. Recently we subtracted the small polydispersity effect and obtained more accurate U(r) with stronger attractions. We did not provide the result here since it is not relevant to the Comment, but we can provide the result if needed.

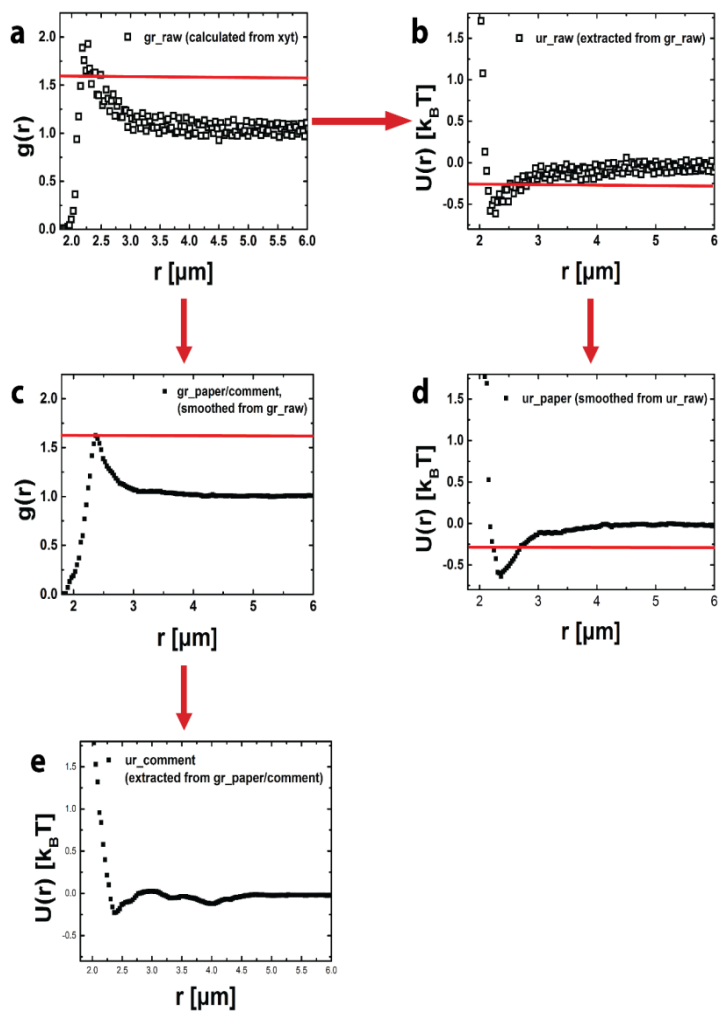

Fig.R1. Smoothing effect in g(r) → U(r) of 30.0°C. (a) Raw g(r). (b) Raw U(r) from the raw g(r). (c) Smoothed g(r) from (a). (d) Smoothed U(r) from (b). (e) U(r) from the smoothed g(r) in (c). (c, d) were reported in ref.1.

## 2. Whether the attraction is tunable (Fig.2a,b, S4 of the Comment)?

**2A (about Fig.2a,b):** In the first version of the Comment submitted in Mar.2016, Cao et al. showed that the attraction is barely tunable in a global heating device (incubator) in Fig.2a. We agree on this. However, all our premelting experiments were performed under an objective heater which heated (1cm)$^2$ circular area in (2cm)$^2$ samples; it's dye-pumping mechanism should be responsible for the tunability of the attraction: the temperature gradient at the perimeter of the $(1cm)^2$ heated area induce the thermophoresis of the dye and pump the dye into the heated region, thus increased the dye concentration and enhanced the attraction in the field of view. Cao agreed that this pumping mechanism exist in the samples, but disagreed that it causes tunable attraction since he did not observed tunable attraction in his experiments in late Jan. 2016 after he learned that our Nature paper was accepted; Now this becomes the key criticism highlighted in the title of the Comment and discussed in the following Section 2B.

The pumping mechanism explains Comment Fig.2b that removing the temperature gradient blurred the crystal-vapor interface. The bulk crystalline structure in Comment Fig.2b was preserved or at least not totally melted during the experimental time, suggesting that attraction still existed after the temperature gradient was removed. More clearly, we observed that particles condensed into large crystals in samples under a uniform global heating, e.g. in a hot room, indicating that the attraction exists although it is barely tunable. Consequently line 56 of the Comment "uniform temperature is not able to stabilize the crystal structure and the crystal-vapor interface" is not correct. The large crystals formed under a uniform temperature also indicates that the temperature gradient is not necessary for crystal formation as described in lines 54-56 in the Supplementary Information of the Comment.

Note: Before the first submission of ref.1 in June 2015, we (including Cao and Liao who were the coauthors at that time) considered the cm sized heated area as 'infinitely large' as many other groups did when they used such objective heater because it is much larger than the $(0.1mm)^2$ field of view in the center of the heated area. Therefore all the measurements were performed with the objective heater. In the first-round review, the referees wanted us to better explain the mechanism of the attraction, although they also agreed that a full understanding on the mechanism is not critical for the publication of ref.1; as long as the tunable attraction is reproducible in equilibrium, it is good to be used as a novel colloidal model system. After the first-round review, Han firstly pointed out that the cm sized heated area may not be considered as 'infinitely large' because the possible thermophoretic effect may pump the dye from the non-heated region into the heated region, thus increase the attraction. We emphasize that this still can be called as tunable attraction at equilibrium because the temperature and the dye concentration in the field of view are uniform. We disagree with the conclusion in the first paragraph of the Comment "We conclude that the melting and premelting reported in their paper are non-equilibrium phenomena caused by non-uniform temperature in the samples of Li *et al*."

**2B (about Fig.S4):** Cao added Fig.S3,S4 in the revised version of the Comment submitted in Apr.2016. **The newly added Fig.S4 showed that the tunable attraction does not exist even under the objective heater, which is the key criticism of the Comment. Since tunable attraction did not exist in Cao's opinion, he interpreted the tunable premelting (Cao also easily observed it) as a result of**

**tiny drift which strength changes with the temperature gradient (Comment Fig.S3).** After we learned this new major criticism (Comment Fig.S4) in Apr. 2016 from Cao, we did various experiments which clearly show that the tunable attraction existed under the objective heater (e.g. Fig.R2). At high temperatures, particles even form clusters, indicating the attraction should be stronger than at least 0.5kT. We need more time to finish quantitative measurements because our software licenses and computing resource are limited. (Cao did the experiment in late Jan. and provided his quantitative g(r) and u(r) in Fig.S4 in Apr.; we just learned his new criticism in Apr. and also need time to do the quantitative analysis.)

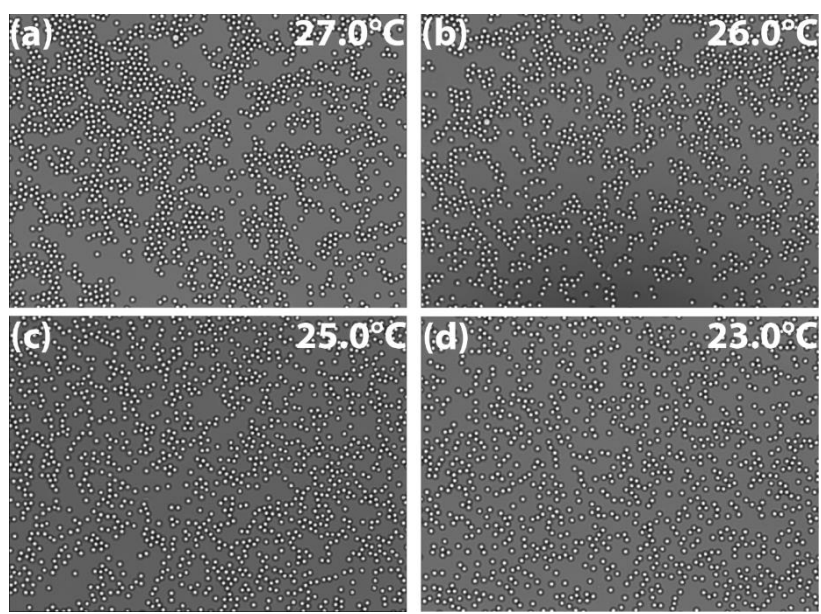

Fig.R2: Raw images of a dilute gas of 1.71 μm PMMA spheres at four different temperature points. Attractions are well tunable at a low dye concentration (10% by volume). At each temperature, we waited 2h for equilibration before taking the video. The room temperature is 22.0°C. [The illumination light is tuned to make the videos at different temperatures have the same brightness.]

**The second author of the Comment, Liao, just finished checking one set of raw videos from Cao and one set of raw videos from Li about g(r) and u(r) measurements for the first time yesterday (i.e. May 11, 2016); She confirmed that Cao's g(r) barely changes with the temperature, while Li's g(r) strongly changes with the temperature.** Li and Cao's measurements were under similar conditions except that Li used a lower dye density, a higher number density of colloidal particles and a smaller size of PMMA spheres.

**Li recently found that the tunability varies under different experimental conditions.** (1) If the dye concentration is too high, the attraction appeared less tunable with more flows (e.g. Fig.R3). **To our knowledge, all Cao's samples looked much darker than Li's good samples** (e.g. Fig.R2). In addition, Cao's g(r) peak heights in Comment Fig.S4a slightly changes at different temperatures, but the corresponding U(r) in Fig.S4b does not change and exhibits some unphysical wiggling, indicating some problems in g(r) → U(r) process. (2) We found that both the ambient temperature T and the heating ΔT affect the system. We used different ambient temperatures ranging from 20°C to 28°C and found that the higher the ambient temperature, the stronger the tiny flow when ΔT's are the same. In ref.1, our ambient temperature is 20°C, lower than 23°C of Cao's experiments. (3) Our recent experiments show that the tunability can be different even under the same dye concentration and ambient temperature. This is probably due to the different ratios of the heated and unheated colloid-dye areas. To make monolayer thin samples, the colloid-dye area needs to be smaller than the whole $(2cm)^2$ sample area; if the colloid-dye area is too small and fully within the $(1cm)^2$ heated area, then the attraction should not be tunable because the unheated area serving as the reservoir of the dye does not exist for pumping. This is consistent with Li's experience that attraction becomes less tunable when he used lots of objective oil because more oil will heated more areas. **This could also explain why Cao did not observe the tunable attraction**. However the heated and unheated areas were not recorded in Li and Cao's experiments because they did not realized that this factor may affect the result. (4) We used 1.7 μm, 2.08 μm and 2.74 μm PMMA spheres and found that larger spheres appeared to have less tunability in attraction, but similar premelting behaviors. Nevertheless, 2.74 μm spheres exhibit tunable attraction in our sample, thus **Cao's non-tunable attraction in Fig.S4 for 2.74 μm spheres is due to some other reasons such as inappropriate high dye density or inappropriate ratio of heated and unheated areas. We will do more experiments to understand why Cao cannot observe the tunable attraction.**

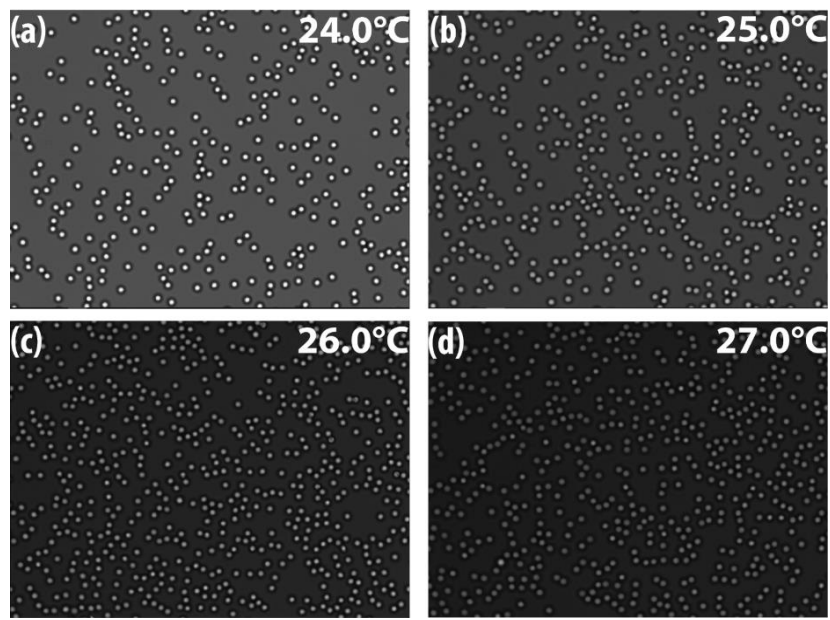

Fig.R3: Raw images of a dilute gas of 2.74 μm PMMA spheres at different temperatures. Attractions are less tunable [but the tunability still exists from the g(r) and u(r) measurement] at a high dye concentration (20% by volume). At each temperature, we waited 2h for equilibration before taking the video. The room

temperature is 23.0°C.

Right before the submission of this Reply on the deadline May 12, 2016, we designed a simple way to unambiguously show that the attraction is tunable. **A quick experiment within one hour shows that attraction is stronger under a higher dye density without any temperature gradient (Fig.R4). Both Cao and us agreed that the pumping effect exists when the temperature gradient is applied, hence pumping more dye into the field of view can enhance the attraction.**

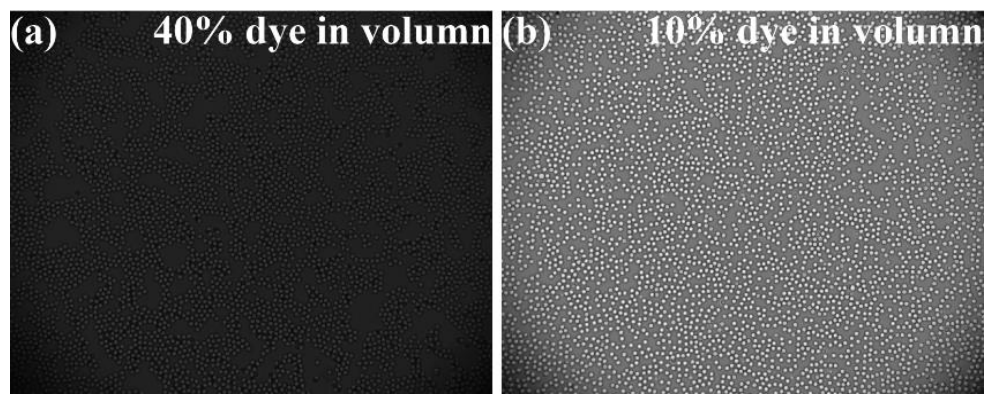

Fig.R4: Raw images of 2.74 μm PMMA spheres in two samples with different dye densities at room temperature 23.0°C without any heating. **It unambiguously shows that attraction depends on the dye density and the temperature gradient is not necessary.**

3. Whether the system is in equilibrium (Fig.2c)?

**Our premelted crystals in ref.1 were in equilibrium because they satisfied the two conditions: (1) temperature is uniform in the (0.1mm)$^2$ field of view because the field of view is tiny and at the exact center of the (1cm)$^2$ uniformly heated area through the thermal contact of the objective oil (thermal equilibrium); (2) no mass flow in the field of view (mechanical equilibrium), e.g. Fig.S8 and movie S7 of ref.1.** Although we did not realize that the pumping effect is responsible for the tunable attraction before the first submission, we still can guarantee that the tunable premelting in our first submission to Nature is under equilibrium since "the two conditions" are satisfied.

We listed the possible non-equilibrium effects criticized by Cao as follows.

**Strong flow:** At high temperature (28°C), the samples occasionally exhibited strong flows which lasted several seconds every ~10 min. At very high temperature (32°C), the strong flow lasted long time and we made use of it for fast epitaxial growth of crystal. (Without making use of such flow, the crystal can also form under an objective heater at lower temperatures or even in a hot room at 27°C without any temperature gradient, but the crystallization rates were much slower.) Our premelting experiment in ref.1 was in the low temperature regime without such large flows. These have been reported in details to referees in Nov. 2015. Referee recommended publication after this clarification.

**Tiny flow (Comment Fig.S3, movies 1, 2):** This flow can be observed by eyes under 240X real time in the movies of the Comment. **Cao did not consider this small effect as one of the four major problems that we should report to editor in Nov.2015** (note: we eventually reported all the major problems and some minor problems to the editors and referees in Nov. 2015). After he observed that the attraction is not tunable even under an objective heater (Fig.S4) in late Jan. 2016, he attributed the tunable premelting as a result of such tiny flow; otherwise the tunable premelting can hardly be explained. Since we found that the attraction is actually tunable under appropriate conditions, it is natural to attribute the premelting as a result of tunable attraction rather than such tiny flow. Moreover, we observed much smaller flows in some premelting samples.

Fig.R5 shows that the 2-layer premelting experiment in ref.1 has **much smaller flows without the monotonic temperature dependence** as shown in Fig.S3, **hence the tiny drift is clear not responsible for premelting.** Note that the tiny flow is too weak to induce hydrodynamic attraction, especially in quasi-2D where the hydrodynamic interaction is short-ranged (comparable to the wall separation of 1.5-diameter of spheres). Since the thermal energy of each particle is much greater than its tiny drift, the tiny drift is unlikely to induce crystallization. Xin Cao posted the Comment in arxiv and then asked Han's Ph.D. advisor, Prof. David Grier, to check his Comment and our ref.1. Prof. Grier commented "**thermophoretic flows may exist, but apparently are too weak to account for the observed phase behavior**."

Using the equation in Supplementary Information of Comment line 62, Cao estimated the equilibration time of the dye is about 10 days for $\Delta T = 10K$. Such theoretical estimation of dye equilibration time needs further tests because the effects of colloidal spheres, glass walls, property of dye etc. are not clear. For example, the thermophoretic mobility of dye molecules is not available. (Thermophoretic mobility was mainly studied for polymers and proteins, which molecular weights are much higher than typical dye molecules.) Even we use the equation in line 62 and Cao's parameters, $\Delta T$ of our premelting experiments range from 2K to 7K (the phase transition regime is < 2k) for different trials of experiments, which corresponds to 50 to 14 days dye-equilibration time. This disagrees with our observations that the premelted liquid layer expanded within one hour after each temperature step, and maintained at a steady state (e.g. the same thickness of the premelted surface liquid phase) in the subsequent several hours at a fixed temperature.

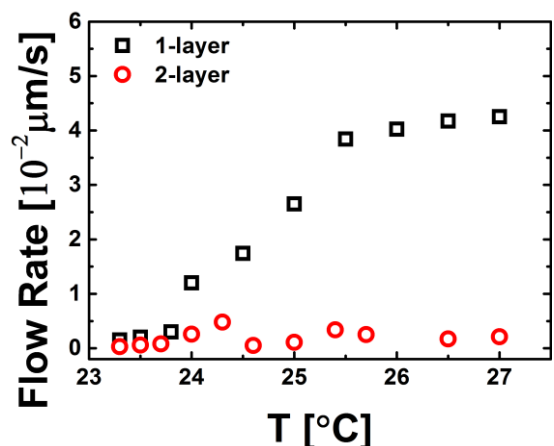

Fig.R5: The tiny flows of the vapor phase in the premelting experiments of monolayer and bilayer crystals in ref.1. The flow rates of the monolayer vapor is comparable to those in Comment Fig.S3, while the flow rates of the bilayer vapor is much less and non-monotonic, i.e. not responsible to the increasing thickness of the premelted liquid.

Fig.2c of the Comment shows that a beam of light optically heated a region smaller than the $(0.1 \text{ mm})^2$ field of view, hence the field of view was not in equilibrium. By contrast, our premelted crystals in ref.1 were in equilibrium because they were in the exact center of a cm-sized uniformly heated area without such local optical heating.

**4. Particle evacuation in Comment fig.2d.**

Fig.2d of the Comment shows that a beam of heating light sometimes can even evacuate instead of aggregate particles when the sample is thick. This exotic phenomenon was firstly observed by Li two years ago. We did not report it in ref.1 because it is not relevant to our premelting experiments which were in thin samples without local optical heating. In thick samples, we and the previous group member, Ziren Wang, **observed such exotic evacuation even without dye** in different colloids although the effect is weaker (due to less light absorption without dye). We observed that (1) in thick samples: (1a) without dye: slight evacuation under strong local optical heating; no effect under weak local optical heating; (1b) with dye: evacuation under strong local optical heating; aggregation under weak local optical heating. (2) In thin samples: (2a) without dye: no effect under local optical heating; (2b) with dye: aggregation under local optical heating. Thick samples appeared to absorb optical heating much more strongly than thin samples and thus may drive the thermophoresis of colloidal spheres. Any dye can enhance the heat absorption and thus enhance this effect, but only the special dye used in our premelting experiment can induce attractions. Cao did not know the above experimental results and did not suggest any possible mechanism of the exotic evacuation.

**Other clarifications:**

Videos S2 and S7 of ref.1 have different vapor densities simply because they are from different trials of premelting experiments. Consequently it does not "violate the basic principle of thermodynamics" as claimed in line 113 of the Comment.

We did not claim that we fully understand the tunable attraction in ref.1. Based on our recent studies in the past half year, we have a better understanding about the system, but still need to do more experiments. The thermophoretic pumping effect far away from the field of view provides a general approach to make other depletion or Casmir-like attractions in colloids tunable as long as the system has thermophoretic effect. We planned to write a research paper about the tunable attraction, but did not expect that Cao *et al.* submitted a Comment and posted it in arxiv. The Comment and Reply are not appropriate because (1) The Comment contains some results firstly discovered by Li in Han's lab and some explanations firstly proposed by Han and Li, and these results will be in our next research paper. (2) The main text of ref.1 has no problem: the premelting was indeed in finely tunable attractive colloidal crystals under equilibrium and the premelting behaviors in ref.1 have been independently reproduced by another student, Tang.

References:
[1] Modes of surface premelting in attractive colloidal crystals, Bo Li, Feng Wang, Di Zhou, Yi Peng, Ran Ni, and Yilong Han, Nature, 531, 485 (2016).

**Appendix II:** Additional results not shown in ref. [1] and Appendix I.

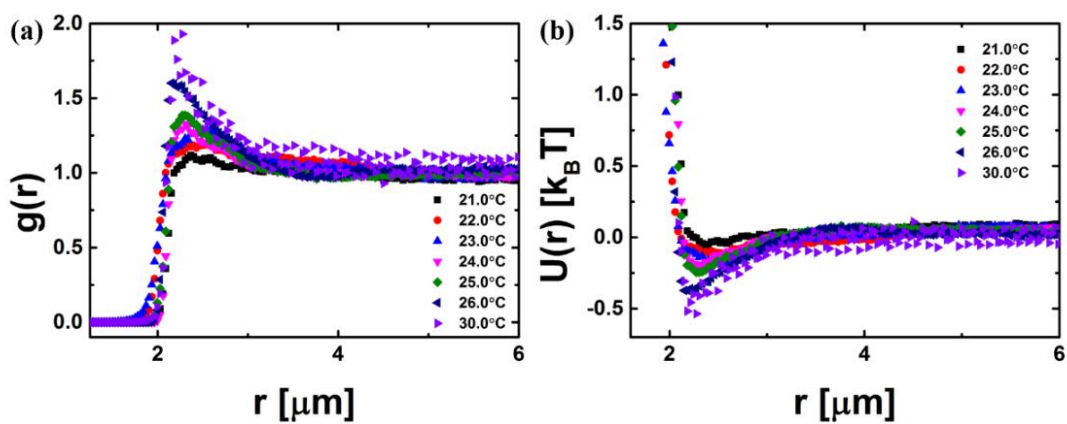

Fig.A1. Unsmoothed (a) g(r) and (b) u(r) corresponding to Fig. S1 and Fig. 1 of ref. [1].

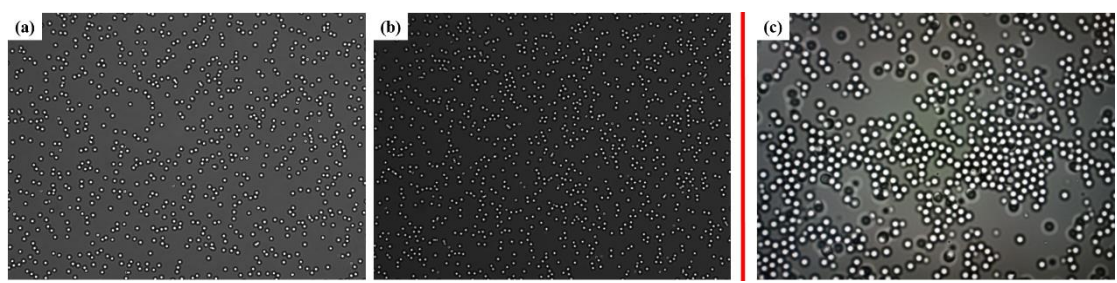

Fig.A2. The tunable attraction observed by Prof. Ziren Wang's group at Chongqing University, reproduced with their permission. After increasing the temperature by 2°C, a homogenous gas in (a) is more clustered in (b). The measured minimum of U(r) is $0.19k_BT$ in (a) and $0.32k_BT$ in (b). The quantified values is in consistancy with the results in Fig. 1 of ref.[1]. (c) is from another experiment performed by a different person. Upon increasing the temperature, a homogesous non-attractive gas (not recored) was condensed into liquid-like clusters shown in (c) within two hours. The field of view was not changed in both experiments. No flow was detected even at 300x real time

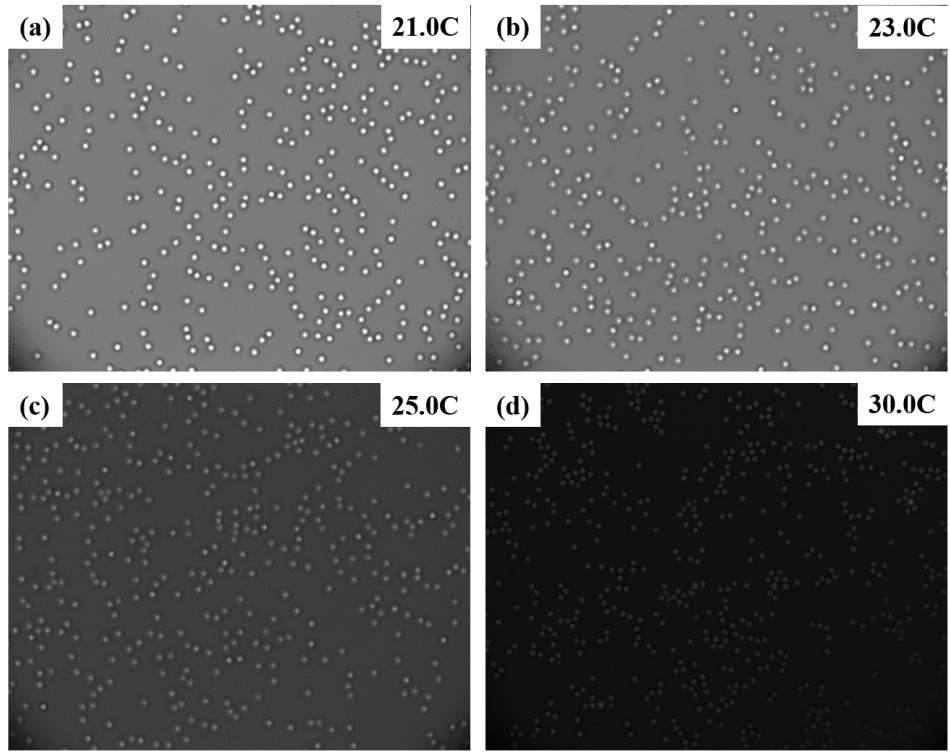

Fig. A3. Raw images of the colloidal vapors used to extract g(r) and u(r) in Figs. 1 and S1 of ref. [1]. The images became darker at higher temperatures, indicating that the dye was pumped into the field of view.

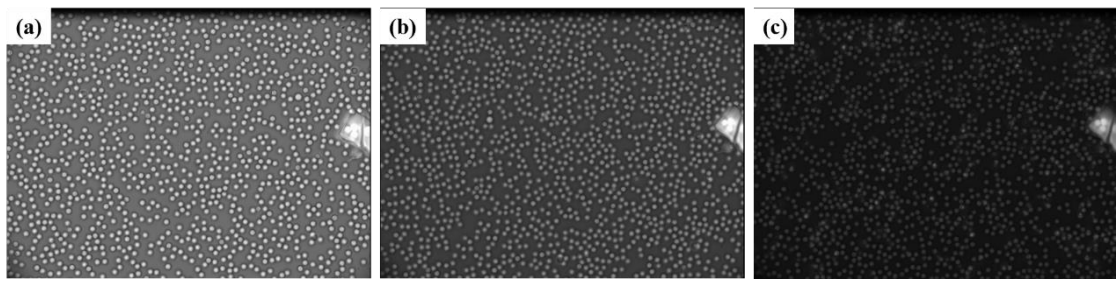

Fig. A4. The raw images at (a) 24.0°C, (b) 26.0°C and (c) 27.0°C show the tunable attraction. The field of view has a stuck object at the right end, which demonstrates that the field of view was unchanged. The illumination light was fixed during the experiment, and the darker images at higher temperatures reflect the dye being pumped into the field of view. The original dye density is 15% in volume and the ratio between the heated and unheated regions is about 0.5 in this experiment.

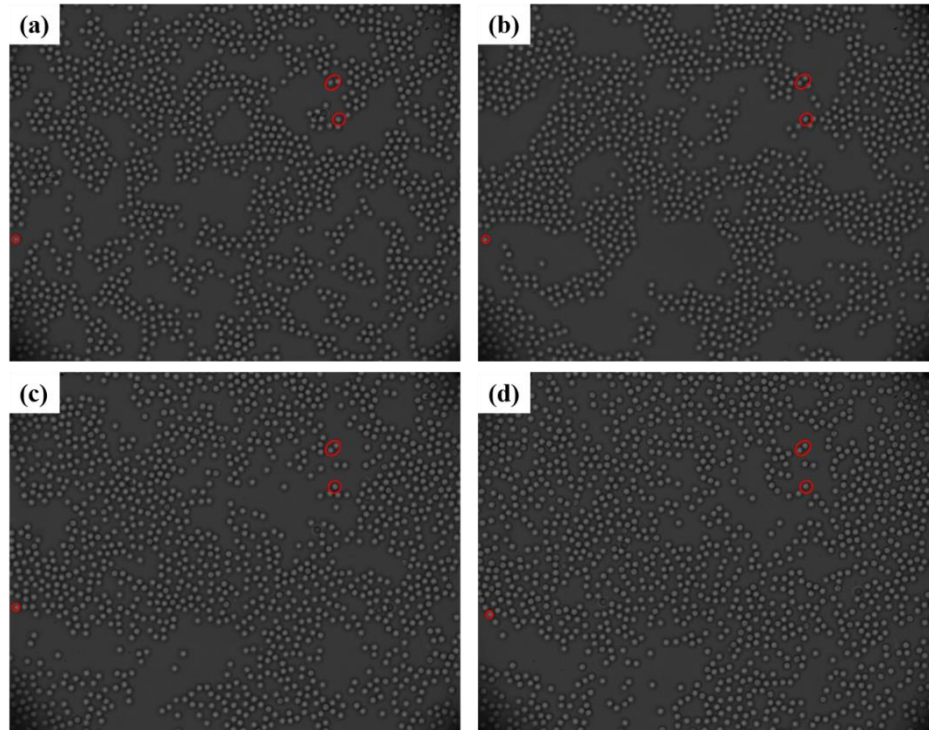

Fig. A5. The raw images at (a) 32.0°C, (b) 30.0°C, (c) 28.0°C and (d) 26.0°C from movie 3 show the tunable attraction. We intentionally chose a field of view with stuck particles labeled with three red circles to rule out the possibility of changing the field of view. Clusters disassociated as the temperature decreased. The illuminance of the light source was changed to maintain a constant image brightness in this experiment.

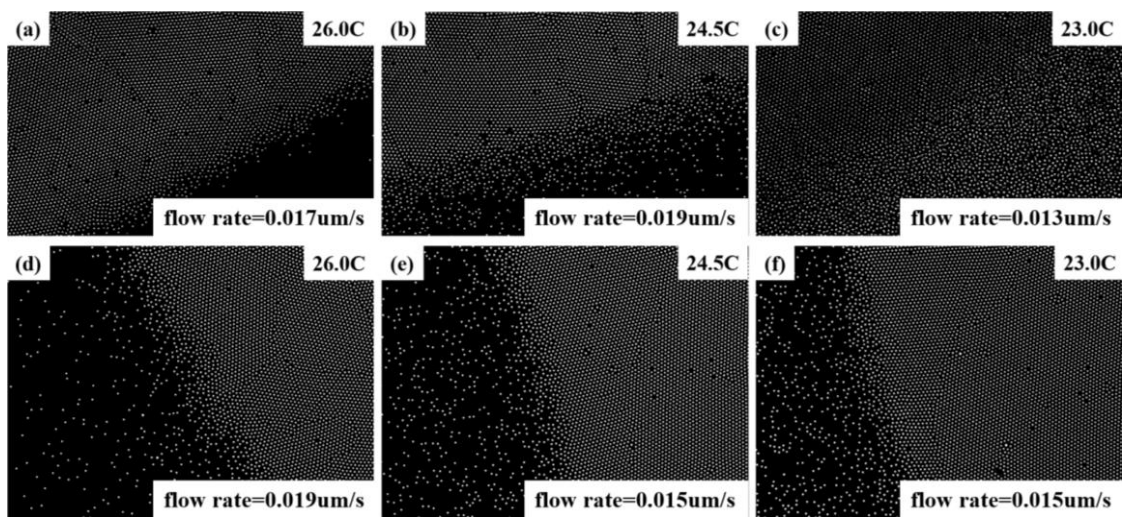

Fig. A6. (a-c) The complete premelting of a bilayer crystal. (d-f) The blocked premelting of a monolayer crystal. (a-c) and (d-f) are in two regions of a single sample with similar flow rates but qualitiatively different premelting behaviors. The flow rates (< 0.02 μm/s) are much lower than Cao's values in ref. [2] and are too low to induce phase-behavior changes. In addition, the three flow rates in (a-c) are non-monotonic and therefore cannot be responsible for the monotonic change in attraction.